\documentclass{cimento}
\pdfoutput=1

\usepackage{graphicx}  
\usepackage{slashed}

\title{Status update: asymptotically safe gravity-matter systems}
\author{Astrid Eichhorn\from{ins:x}}
\instlist{\inst{ins:x} CP3-Origins, University of Southern Denmark, Campusvej 55, 5230 Odense M, Denmark}

\usepackage{amsmath}

\newcommand{\bea}{\begin{eqnarray}}
\newcommand{\eea}{\end{eqnarray}}
\newcommand{\be}{\begin{equation}}
\newcommand{\ee}{\end{equation}}

\begin{document}

\maketitle

\begin{abstract}
The field of asymptotically safe matter-gravity systems is maturing from the study of simple toy models to the exploration of sectors of the Standard Model and beyond. This status update reviews the current state of the art and points out open questions and future perspectives.
\end{abstract}

\section{Introduction: A symmetry principle for the fundamental description of gravity and matter}
The Standard Model of particle physics (SM) and General Relativity (GR) are extremely successful effective field theories, describing the outcome of a multitude of experiments. Both can be quantized perturbatively, but break down at large energy scales. In the SM, this breakdown is encoded in transplanckian Landau poles, which prevent an ultraviolet (UV) extension beyond the scale of the Landau poles and signal a triviality problem. In GR, this breakdown is encoded in the infinitely many counterterms, which cause a breakdown of predictivity, because each counterterm comes with a free parameter. 
As a consequence, we only understand the gravitational interaction of elementary particles at subplanckian energies, where it is extremely weak and thus negligible. At the Planck scale, where the gravitational interaction between elementary particles is expected to be non-negligible, the combined effective field theory of the SM and GR breaks down.

This problem is often approached by conjecturing novel symmetries. Many symmetry principles, such as local and/or global symmetries in the SM and GR as well as Beyond SM (BSM) theories, are already present in a classical theory. However, we are looking for a quantum theory of all fundamental forces and elementary particles. Thus, we are led to ask, whether there is a symmetry principle that is inherent to the quantum nature of those forces and particles. The answer is given by the asymptotic safety framework: Quantum scale symmetry, i.e., scale symmetry generated by quantum fluctuations, enables a UV extension of effective field theories and imposes predictivity on effective field theories. Thus, both problems of the combined effective field theory of the SM and GR can in principle be addressed by quantum scale symmetry.

Quantum scale symmetry corresponds to an interacting fixed point of the Renormalization Group (RG). The RG encodes how the dynamics of a theory change, as virtual quantum fluctuations are integrated over in the path integral. For a generic set of fields and interactions, the RG flow is non-vanishing, i.e., quantum fluctuations typically generate a scale dependence in the theory. Scale symmetry is achieved at fixed points of the RG. At free (non-interacting) fixed points, the effect of quantum fluctuations is switched off; thus the existence of free fixed points is automatic. In contrast, at interacting fixed points, quantum fluctuations are present and balance out in a non-trivial way to generate scale symmetry. 

Scale symmetry can be detected in the dimensionless counterparts $g_i(k)$ of all couplings $\bar{g}_i$ of the theory; where $i$ labels all couplings and typically $i \in [1,\infty)$ and where $k$ is an energy/momentum scale. In these dimensionless couplings, the explicit presence of a scale is removed and scale symmetry implies $g_i(k) = \rm const$.

A given fixed point at $g_i=g_{i\, \ast}$ can be realized either in the UV or the infrared (IR), depending on the choice of RG trajectory: given the values of all $g_i$ at some finite scale $k_0$, $g_i(k_0) = g_{i,0}$, we can follow the RG flow to the IR to determine whether the fixed point is reached in the IR. A given fixed point is reached in the IR, if the initial conditions $g_{i\,0}(k_0)$ lie in its IR critical hypersurface. From the UV to the IR is the natural direction of the RG flow, agreeing with its interpretation as a form of coarse graining. Purely mathematically, the RG flow can be inverted to follow it into the UV and determine whether the ``initial conditions" $g_(k_0)= g_{i,0}$ can be reached from the fixed point $g_{i\, \ast}$ as the UV fixed point. This occurs, if the initial conditions lie in the fixed point's UV critical surface. A given fixed point is therefore, unless it has a zero-dimensional UV or IR critical surface, not a priori a UV or IR fixed point -- it depends on the trajectory, that is selected.\\
The UV critical surface is spanned by the relevant directions, i.e., (superpositions of) couplings\footnote{The relevant and irrelevant directions at an interacting fixed point are often superpositions of couplings. If this is the case, an adapted basis in the space of couplings can be chosen.}, along which quantum fluctuations drive the dynamics away from quantum scale symmetry. The IR critical hypersurface is spanned by the irrelevant directions, i.e., (superposition of) couplings, along which quantum fluctuations drive the dynamics towards quantum scale symmetry. \\
Relevant and irrelevant directions are associated to positive and negative critical exponents. Critical exponents parameterize the linearized flow about a fixed point:
\be
g_i(k) = g_{i\, \ast} + \sum_j c_j \, V^j_i\left(\frac{k}{k_0} \right)^{-\theta_j},\label{eq:linflow}
\ee
where $V^j$ is the $j$th eigenvector of the stability matrix that is built from derivatives of the beta functions $\beta_{g_i} = k\partial_k\, g_i(k)$:
\be
\left(\frac{\partial \beta_{g_i}}{\partial g_j} \right)\Big|_{\vec{g} = \vec{g}_{\ast}} V^j = - \theta_j \, V^j.
\ee
The eigenvalues of the stability matrix are multiplied by an additional negative sign to obtain the critical exponents $\theta_i$. In this way, the critical exponents at the free fixed point correspond to the canonical dimensionality of the couplings $\bar{g}_i$. At an interacting fixed point, interactions  at the fixed point shift the scaling spectrum (the set of critical exponents) away from their canonical values.

Even if the fundamental interactions are scale invariant, there must be a scale $k_{\rm tr}$ at which the RG flow departs from a UV fixed point, because elementary particles and their fundamental interactions exhibit scale-dependence at low energies (low $k$). Thus, the constants of integration, $c_j$ in Eq.~\eqref{eq:linflow}, determine the low-energy values of the couplings:
Relevant couplings are not theoretically restricted in their departure from scale symmetry: They can depart from scale symmetry at any scale and a range of values for these couplings is compatible with scale symmetry in the UV. If there are several relevant couplings in the theory, then there are (largely independent) IR ranges for each one of them, parameterized by different sets of choice of the corresponding $c_j$. Thus, the values of $c_j$ for relevant directions can only be determined from experimental input, but are free parameters of the theory.

We now consider the predictions of coupling values that arise from a UV fixed point and are connected to the irrelevant couplings.
If the linearized flow about the fixed point was an exact description of the flow even further away from the point $g_{i\, \ast}$, then irrelevant couplings would have to satisfy $g_i(k) = g_{i\, \ast}$ for all $k$. Hence, their values would be fixed at all scales by the requirement of UV scale symmetry. The deviation of relevant couplings from the fixed-point value, which occurs below the transition scale $k$, would leave the irrelevant couplings untouched.
However, further away from the point $g_{i\, \ast}$, the critical surface exhibits curvature. Thus, when relevant couplings depart from the fixed-point values, they pull the irrelevant couplings with them. The values of irrelevant couplings are still determined at all scales, because even in the deep IR, far away from the fixed-point regime, they remain fully determined by the relevant couplings; the $c_j$ for irrelevant directions are not free parameters.

\section{Observables and observations}
Irrelevant couplings are predicted (at all scales $k$) by requiring quantum scale symmetry in the UV. These predictions are testable, because couplings enter observables, such as scattering cross sections, masses etc. These predictions concern experimentally accessible (e.g., by the LHC) scales. In most cases, the relevant experimental measurements have already been made. Thus, the data to test predictions resulting from an asymptotically safe UV extension of the SM with gravity, already exists -- and just waits for us to test quantum gravity with it \footnote{The same reasoning applies in other quantum gravity approaches: They all have a landscape (matter theories compatible with a UV completion within the given quantum-gravity approach) and a swampland (matter theories compatible with a UV completion within the given quantum-gravity approach). The size of swampland and landscape might be very different from approach to approach, thus some approaches might be much easier to rule out than others. In practise, most is known or conjectured about the swampland/landscape in string theory and asymptotic safety (using functional Renormalization Group tools -- a few results find support from lattice approaches, such as the existence of light fermions \cite{Eichhorn:2011pc} in \cite{Catterall:2018dns}).}.

Such tests of quantum gravity are based on low-energy observables. This circumvents a tricky problem in quantum gravity, namely that (formal) observables have to be non-local in the absence of a preferred background (and thus rather difficult to access in practise -- calling into question the practise of calling them ``observables" without the clarifying descriptor ``formal"), because at low energies, either a (near) flat background emerges dynamically from a quantum gravity theory, or that theory is ruled out because it cannot reproduce the large-scale spacetime structure in our universe. On a flat background, the standard, observationally accessible particle-physics observables indeed emerge from the formal observables of background-independent quantum gravity.

\section{Effective and fundamental asymptotic safety and their phenomenology}
An RG fixed point with at least one relevant and at least one irrelevant direction can be phenomenologically relevant in two different ways, assuming that there is at least one trajectory emanating from the fixed point to a regime where predictions agree with observations: First, it can be a true UV fixed point at which one can formally take $k \rightarrow \infty$. Towards the IR, the flow then departs from the fixed point along its relevant direction(s) at $k=k_{\rm tr}$. Second, it can be an intermediate fixed point, at which one cannot take the formal $k \rightarrow \infty$ limit. Instead, beyond $k_{\rm UV}$, a different, microscopic theory is realized (one can in principle think of string theory, loop quantum gravity or even causal sets here). The microscopic theory provides initial conditions for the RG flow in its effective description, i.e., it determines the values of all couplings at $k_{\rm UV}$. If those lie on the IR critical surface of the fixed point, then the flow passes very close to the fixed point on its way to the IR. The fixed point generates universality, in that different initial conditions on its IR critical surface lead to very similar low-energy values of the couplings. For the irrelevant directions of the fixed point, the low-energy values lie very close to the predictions generated by a true fixed-point trajectory.\\
This idea has first been put forward in \cite{Percacci:2010af}, expanded in \cite{deAlwis:2019aud,Held:2019vmi} and applied to the string-theory context in \cite{deAlwis:2019aud,Basile:2021euh,Basile:2021krk,Basile:2021krr}. A quantitative measure of predictivity in effective asymptotic safety has been developed in \cite{Held:2020kze}.

\section{Key basics of functional Renormalization Group flows}
Here, the question whether or not gravity on its own is asymptotically safe or not is left aside. This question is irrelevant for a description of quantum gravity in our universe \footnote{We work under the assumption that matter is at least partly non-emergent, i.e., that a quantum field theory of the metric, if asymptotically safe, would not give rise to the matter and force fields of the SM as emergent, collective, low-energy degrees of freedom.}.
Most evidence for asymptotic safety in gravity-matter systems comes from functional Renormalization Group techniques \cite{Wetterich:1992yh,Morris:1993qb}, reviewed in \cite{Dupuis:2020fhh}.
Reviews that focus mostly on the asymptotic-safety paradigm for gravity can be found in \cite{Reuter:2012id,Pereira:2019dbn,Pawlowski:2020qer}, text books in \cite{Percacci:2017fkn,Reuter:2019byg}. The interplay with matter is reviewed in \cite{Eichhorn:2017egq,Eichhorn:2018yfc} and recent lecture notes on the asymptotic-safety paradigm are \cite{Eichhorn:2020mte,Reichert:2020mja}. Critical reflections of the current state of the art can be found in \cite{Donoghue:2019clr,Bonanno:2020bil}. 

\begin{figure}[!t]
\includegraphics[width=\linewidth,clip=true,trim=0cm 2cm 0cm 11cm]{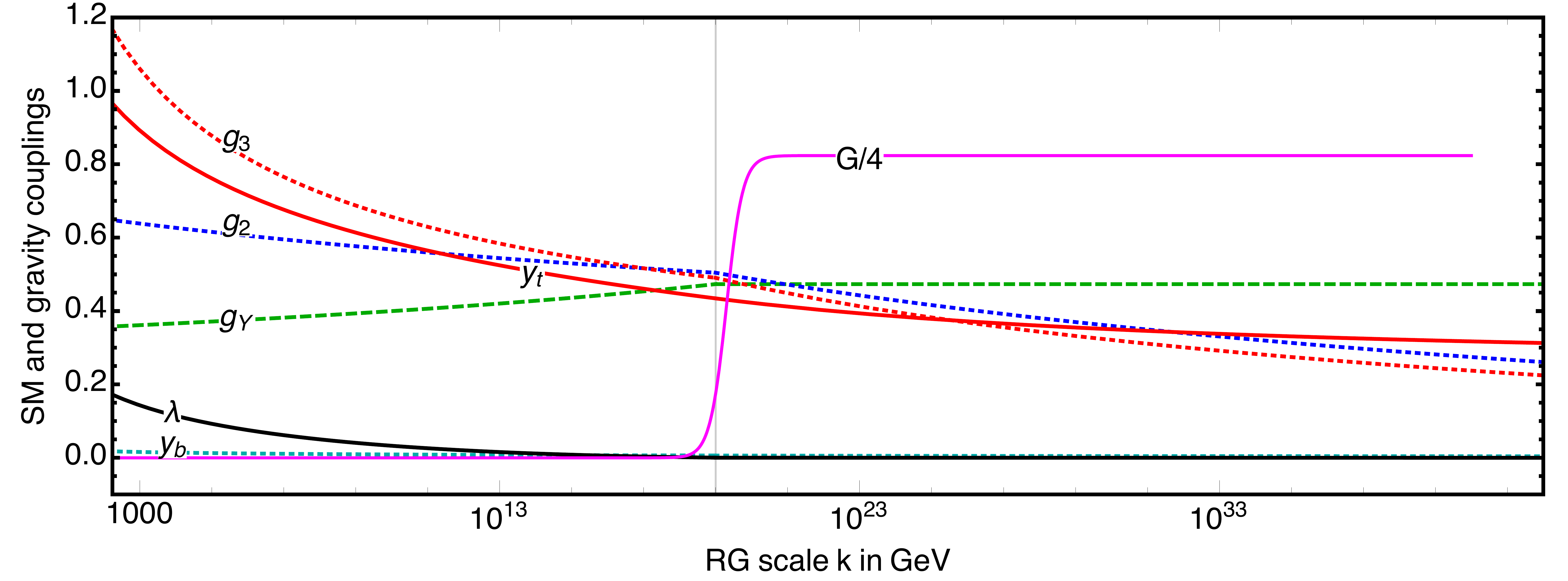}
\caption{\label{fig:flow} An example of a gravity-matter RG flow, following \cite{Eichhorn:2018whv}: Beyond the Planck scale, couplings are either constant (asymptotically safe), or grow away from an asymptotically free fixed point. At highly transplanckian scales, a fixed point is realized which is interacting in a subset of couplings and free in others. The Newton coupling $G$ is a relevant coupling and can be chosen to depart from its fixed-point value at any scale. To reach the correct low-energy value, the scale of departure is chosen to be the Planck scale. As soon as $G$ departs from its fixed-point value, its canonical scaling $G \sim k^2$ drives it to very small values. Thus, gravity fluctuations decouple dynamically at the Planck scale. Next, all irrelevant matter couplings that are held at interacting fixed-point values by gravity, start to run. For these couplings, asymptotic safety provides a unique initial condition at the Planck scale. In contrast, for relevant couplings, a range of initial conditions is available at the Planck scale.
At $k \rightarrow 0$, couplings approach low-energy values that can be compared to experiment. In the subplanckian regime, $k < M_{\rm Planck}$, the RG flow of the marginal Standard Model couplings (strong gauge coupling $g_3$, SU(2) gauge coupling $g_2$, hypercharge coupling $g_Y$, top Yukawa coupling $y_t$, bottom Yukawa coupling $y_b$ and Higgs quartic coupling $\lambda$) exhibit a flow closely resembling the perturbative RG flow in the SM. The most significant difference is that $\lambda>0$ at all scales, resulting in a Higgs mass that lies above the measured value.}
\end{figure}

The functional RG approach provides a differential equation for the scale derivative of the effective dynamics, based on an IR cutoff in the generating functional. This equation can be viewed as a compact summary of all beta functions of the theory. It allows to search for a fixed point, $\beta_{g_i} = 0$. In a second step, once a fixed point is found, the beta functions can be integrated from the corresponding initial condition $g_i(k = k_{\rm UV}\gg k_{\rm tr}) = g_{i\, \ast}$ to obtain the corresponding predictions $g_i(k \rightarrow 0)$. The RG flow of gravity-matter systems has three regimes, see Fig.~\ref{fig:flow}.\\
The $k$-dependence of couplings is not directly physical, see \cite{Donoghue:2019clr,Bonanno:2020bil}, because a given physical system does not typically feature an IR cutoff that resembles the cutoff in the setup of the functional RG. The limit $k \rightarrow 0$ is physical, because it corresponds to the limit in which all quantum fluctuations have been integrated over\footnote{A \emph{physical} scale dependence of couplings is present in this limit, because appropriately defined couplings depend on physical scales, such as momenta, curvature scales, etc.}. Therefore, the predictions that arise for the low-energy values of couplings are physical predictions relevant for a comparison with experiment. 

The beta functions of the functional RG approach are related to those in a perturbative setting as follows: For canonically marginal (i.e., dimensionless) couplings, the one-loop coefficients are universal in that they agree between all schemes. This holds for those one-loop coefficients that  themselves  depend on marginal couplings. Beyond one loop, the universality of beta-functions of marginal couplings is lost (beyond two loops if one works in a mass-independent scheme). For canonically non-marginal couplings, already the one-loop coefficient is non-universal.\\ 
Thus, physical observables -- many of which can be calculated using a combination of different beta functions -- arise from different ``building blocks" in different schemes.
Another universal piece of information encoded in beta functions is the existence of a fixed point. A final universal piece of information is the set of critical exponents, and thus the predictivity of a fixed point.

As a key open question, the functional RG approach to asymptotic safety is based on Euclidean signature. It is a main assumption of the approach, that Euclidean-signature results have some relevance for Lorentzian quantum gravity. In the deeply non-perturbative regime of quantum gravity, this is a strong assumption. However, \cite{Eichhorn:2018akn,Eichhorn:2018ydy,Eichhorn:2018nda,Eichhorn:2020sbo} find indications that asymptotically safe gravity-matter systems exhibit near-perturbative behavior and \cite{Eichhorn:2016esv,Christiansen:2017gtg,Eichhorn:2017eht,deBrito:2021pyi,Eichhorn:2021qet} find indications that strongly-coupled gravity-matter systems are not asymptotically safe. We conjecture, that such near-perturbative behavior implies that a near-flat background emerges dynamically, already at scales around the Planck scale. Around such a background, an analytical continuation from Euclidean to Lorentzian signature is possible (for suitable behavior of the graviton propagator \cite{Bosma:2019aiu,Platania:2020knd,Bonanno:2021squ}), and indeed a Lorentzian propagator has recently been calculated \cite{Fehre:2021eob}. Therefore we conjecture that results on near-perturbative, Euclidean, asymptotically safe gravity-matter systems do in fact have relevance for our Lorentzian universe.

\section{Unavoidable interactions}
When gravity is interacting, one expects that it will mediate effective matter interactions - simply because gravity fluctuations cannot be decoupled from matter. Indeed,  certain matter self-interactions and non-minimal matter-gravity couplings are unavoidable at an asymptotically safe fixed point  \cite{Eichhorn:2011pc, Eichhorn:2012va, Eichhorn:2016esv,Christiansen:2017gtg,Eichhorn:2017eht,Eichhorn:2017sok,Eichhorn:2018nda,deBrito:2021pyi,Laporte:2021kyp,Eichhorn:2021qet}. These interactions are determined by the global symmetries of the kinetic terms of the matter fields. These kinetic terms give rise to matter-gravity vertices and act as ``seeds" from which gravity fluctuations generate matter interactions. The generated matter interactions are symmetric under the global symmetries of the kinetic terms. This observation has been explained based on the diagrammatic structure of the flow equation in \cite{Eichhorn:2020mte}, and has been shown more formally for shift-symmetry in scalar-gravity models in \cite{Laporte:2021kyp}. In \cite{Ali:2020znq} it has additionally been shown that the full global symmetry of the kinetic term for scalars is not broken to a discrete symmetry by gravity fluctuations.
In \cite{Eichhorn:2020mte}, the result has been discussed in the light of the folklore theorem that quantum gravity must break all global matter symmetries. In short, either asymptotic safety is a counterexample (which might be related to the possibility of black-hole remnants), or the Euclidean nature of functional RG calculations implies that symmetry-breaking black-hole configurations are not properly accounted for in the path integral. Topological fluctuations might also play a role in this context -- although it is not established whether or not topology actually fluctuates in quantum gravity -- and \cite{Hamada:2020mug} has recently argued that these would lead to the breaking of chiral symmetry for fermions.

The induced interactions are not expected to be accessible to direct measurements. This is, because interactions that respect the global symmetries of the kinetic terms of various matter fields are all canonically higher order interactions. Thus, even if they are finite in the asymptotically safe fixed-point regime, they are quickly driven to zero in the subplanckian regime of the flow, where gravity fluctuations decouple and we assume that perturbation theory becomes a good approximation. Nevertheless, these couplings give rise to important constraints on the asymptotically safe fixed-point regime, discussed below. 

\subsubsection{The weak-gravity bound}\label{subsec:WGB}
The first constraint on the fixed-point regime arises from scalar self-interactions, vector self-interactions, and fermion-scalar interactions. Interactions like $\left(g^{\mu\nu}\partial_{\mu}\phi\partial_{\nu}\phi\right)^2$, $\left(g^{\mu \kappa}g^{\nu\lambda}F_{\mu\nu}F_{\kappa \lambda}\right)^2$ and $\bar{\psi}\slashed{\nabla}\psi \,g^{\mu\nu}\partial_{\mu}\phi\partial_{\nu}\phi$ (with $\phi$ a scalar, $\psi$ a fermion and $F_{\mu\nu}$ the field-strength tensor) cannot have vanishing couplings under the impact of gravity. Thus, the existence of a fixed point for these couplings is a crucial question, unlike for couplings which can remain zero under the impact of gravity. In \cite{Eichhorn:2016esv,Christiansen:2017gtg,deBrito:2021pyi}, it was found that once gravity fluctuations exceed a critical strength, they push the fixed-point values for these couplings into the complex plane. At that point, an asymptotically safe gravity-matter fixed point can no longer exist. In turn, the gravitational couplings are constrained to be sufficiently weak. The bound beyond which they push matter couplings into the complex plane is the \emph{weak-gravity bound}, which bounds the regime in which gravity is sufficiently weakly coupled to allow for \emph{minimally interacting} fixed points in the matter sector.

\begin{figure}
\begin{minipage}{0.45\linewidth}
\includegraphics[width=\linewidth]{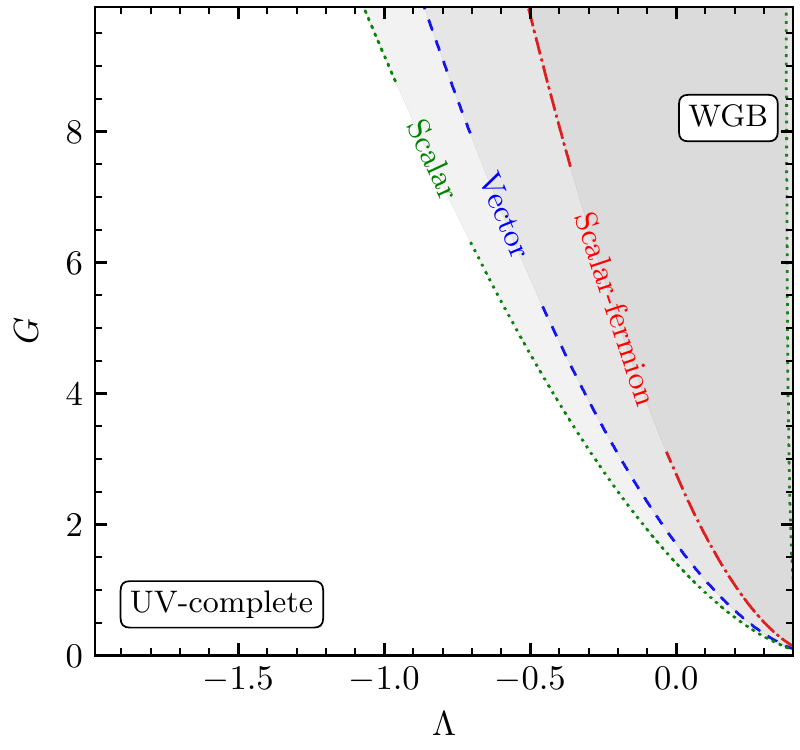}
\end{minipage}
\quad \quad
\begin{minipage}{0.45\linewidth}
\caption{\label{fig:WGB} Comparison of the weak-gravity bound in the plane spanned by the dimensionless Newton coupling $G$ and dimensionless cosmological constant $\Lambda$. The region above and to the right of the green dotted (blue dashed) line (in grey; labelled by "WGB") is excluded when scalars (gauge fields) are present; the region above the dot-dashed line is excluded when scalars and fermions are present.\newline\\}
\end{minipage}
\end{figure}

The weak-gravity bounds that arise from different matter sectors delineate a qualitatively similar region in the gravitational parameter space \cite{Schiffer:2021gwl}, cf.~Fig.~\ref{fig:WGB}. Furthermore, the boundary depends only weakly on the number of matter fields \cite{deBrito:2021pyi,Eichhorn:2021qet}.
Thus, the different matter sectors distinguish a preferred and an excluded region in the gravitational parameter space. Gravitational fixed-point values lie close to the boundary of the excluded region for purely gravitational systems (where it depends on the choice of truncation, whether the gravitational fixed point ends up beyond or below the boundary \cite{Eichhorn:2012va,deBrito:2021pyi,Laporte:2021kyp}), but move away from this region, when appropriate matter fields (with spin 1/2 or spin 1) are added to the system \cite{deBrito:2021pyi}.

\subsubsection{Light fermions}
The lightness of fermions compared to the Planck scale is one of the remarkable aspects of the SM, giving rise to fundamental and composite particles for which the gravitational interaction is completely negligible at energies close to the fermionic mass scales. This is due to chiral symmetry, which implies that left-handed and right-handed fermions transform separately and which would be violated by an explicit mass term. In the SM, it is broken spontaneously by the electroweak symmetry breaking and in QCD. In the SM, these two effects give rise to the observed masses of fermions and their bound states.\\
If another interaction would break chiral symmetry at higher scales, it would generate correspondingly more massive bound states and thus be incompatible with observations. Thus, diagnosing whether quantum gravity fluctuations lead to the breaking of chiral symmetry and generation of bound states is an important test.\\
Chiral symmetry-breaking is linked to divergences in four-fermion interactions: when these diverge, a corresponding bound state forms. In \cite{Eichhorn:2011pc}, it was shown that gravity prevents this mechanism from occurring, despite inducing four-fermion interactions. No matter how strong gravity fluctuations become, they always lead to finite values of four-fermion interactions \cite{Meibohm:2016mkp,Eichhorn:2017eht}. This is because such gravitational contributions that would lead to a loss of fixed points (and thus a divergence), exist (just like they do for the interactions that feature a weak-gravity bound), but are counteracted by gravitational interactions that produce a change in scaling dimension of these operators.\\
The situation changes once an Abelian gauge field is added under which the fermions are charged. Then, gauge field fluctuations can be strong enough to overcome gravitational fluctuations, and lead to chiral symmetry breaking at the Planck scale. This mechanism can be prevented, if gauge field fluctuations cannot become strong enough. In turn, the fixed-point value for the gauge coupling, and thus the strength of gauge field fluctuations, decreases with the number of charged fermion species in the system. As a consequence, light charged fermions can exist only above a critical number of charged fermions. A first estimate of this critical number in \cite{deBrito:2020dta} is approximately 3. Such a mechanism might in fact explain why several generations of fermions exist. \\
Additionally, upper bounds on the number of fermion species arise, if gravitational catalysis due to background spacetime curvature becomes relevant \cite{Gies:2018jnv,Gies:2021upb}. Finally, \cite{Hamada:2020mug} has argued that topological fluctuations of spacetime provide an additional symmetry-breaking effect.

\section{Status of the asymptotically safe Standard Model with gravity}
\subsection{Parameterized approach: Available universality classes}
In the literature, two different ways of reporting the effect of gravity on matter are used: In the first, gravitational fixed-point values within a given truncation are used and quantitative effects of gravity on matter are reported. In the second, gravitational fixed-point values are left as free parameters. In this way, an understanding of possible quantum-gravity effects throughout the entire parameter space can be developed (and the first approach can be taken at a later stage). This has the added advantage that one can straightforwardly generalize from a parameterization of the gravitational parameter space to a parameterization of more general new physics: gravity contributes to the flow of canonically marginal matter couplings through anomalous scaling terms, e.g.,
\bea
\beta_{g_Y} &=& - f_g\, g_Y+\frac{41}{6\cdot 16 \pi^2}g_y^3+...\label{eq:gy}\\
\beta_{y_{b/t}}&=& - f_y\, y_{b/t}+\frac{y_{b/t}}{16\pi^2}\left(\frac{3}{2}y_{b/t}^2+\frac{9}{2}y_{t/b}^2 -3\left(\frac{1}{36}+Y_{t/b}^2\right)g_Y^2\right)+... \label{eq:tb},
\eea 
where $g_Y$ is the Abelian hypercharge coupling and $y_{b/t}$ are the bottom and top quark Yukawa coupling, respectively, which carry hypercharges $Y_t=2/3$ and $Y_b=-1/3$. 
$f_g$ parameterizes the gravitational contribution to the Abelian gauge coupling (which is the same for the other two gauge couplings in the SM) and $f_y$ parameterizes the gravitational contribution to the Yukawa couplings (that is the same for all Yukawa couplings in the SM). Both $f_g$ and $f_y$ depend on the gravitational fixed-point values and explicit expressions can be found in \cite{Oda:2015sma,Eichhorn:2016esv,Eichhorn:2017eht} for $f_y$ and \cite{Harst:2011zx,Eichhorn:2017lry} for $f_g$; both are constant beyond the Planck scale and vanish quickly below the Planck scale due to the scaling $G \sim k^2$ of the Newton coupling that both are proportional to.\\
From Eq.~\eqref{eq:gy} and \eqref{eq:tb}, we see that $f_g<0$ and $f_y<0$ are phenomenologically problematic: for these values, $g_Y$ and $y_{b/t}$ only have IR attractive fixed points at zero. Thus, these couplings vanish in the UV, down to the Planck scale. Below, they remain zero because symmetries forbid their generation, once they are set to zero.\\
Further, from Eq.~\eqref{eq:gy} and \eqref{eq:tb} we see that for $f_g>0$ and $f_y>0$, the free fixed points in the matter couplings are IR repulsive, allowing to reach nonvanishing IR values of these couplings.\\
Finally, from Eq.~\eqref{eq:gy} and \eqref{eq:tb} we see that there are interacting fixed points for $f_g>0$ and $f_y>0$. These are IR attractive, and thus provide not just a non-vanishing value of these couplings in the UV, but also a unique, nonvanishing value in the IR. From the combination of Eq.~\eqref{eq:gy} and \eqref{eq:tb}, we also see that a nonzero fixed-point value $g_{Y\, \ast}>0$ is necessary to generate a difference between the predicted value for top and bottom Yukawa. In \cite{Eichhorn:2018whv} it was shown that Eq.~\eqref{eq:gy} and Eq.~\eqref{eq:tb}, supplemented by the beta functions for the non-Abelian gauge couplings, which remain asymptotically free in the presence of gravity \cite{Daum:2009dn,Folkerts:2011jz}, values for $f_y$ and $f_g$ exist, for which $g_Y, \, y_t$ and $y_b$ take on IR values in the vicinity of their SM values.

\subsection{Asymptotically free quark sector}
Going beyond the third generation of quarks in Eq.~\eqref{eq:gy} and \eqref{eq:tb}, CKM mixing has to be taken into account. It turns out that the CKM matrix elements run very slowly, so that to see the effect of their running, we have to consider immense ranges of scales \cite{Alkofer:2020vtb}. One can have the following attitude to this range of scales: i) if asymptotic safety is fundamental, then a QFT description in principle makes sense to arbitrarily high scales, thus the huge range of scales may be unusual, but not nonsensical; ii) if asymptotic safety is effective, then such huge ranges do not make sense and then CKM running can be neglected for practical purposes.\\
In \cite{Alkofer:2020vtb}, it was found that under the impact of gravity, the gravity-generated, asymptotically free fixed point that exists for $f_g>0$ and $f_y>0$, can be connected to IR values for gauge and Yukawa couplings and CKM matrix elements which are those of the SM. On the way from the UV fixed point to the IR, the flow passes in the vicinity of interacting fixed points, which leave their imprint on the IR predictions.\\
The full situation including leptons is currently an open question; the third generation including leptons and a neutrino Yukawa coupling was investigated in \cite{Held:2019vmi}.
\subsection{Preferred region in parameter space and gravity-matter interplay}
For the gauge coupling, $f_g>0$ is realized across all values of $G$ and $\Lambda<0.5$ (the latter bounds the basin of attraction of the gravitational fixed point) \cite{Daum:2009dn,Folkerts:2011jz,Christiansen:2017cxa,Eichhorn:2017lry}. A similar statement is true for $f_{\lambda}$, which is the analogous quantity to $f_g$ and $f_y$, but for the Higgs quartic coupling, see \cite{Narain:2009fy,Eichhorn:2017als,Pawlowski:2018ixd,Eichhorn:2020sbo}. 

For $f_y$, the situation is more intricate, because it is not positive everywhere in the gravitational parameter space. Thus, phenomenological viability singles out the region, where $f_y>0$ holds, see \cite{Oda:2015sma,Eichhorn:2016esv,Eichhorn:2017eht}. Specifically, the region that is excluded has significant overlap with the strong-gravity-regime, which is excluded by the weak-gravity bound. Thus, two independent mechanisms rule out (partially overlapping) parts of the strong-gravity region.

This is a further indication for the near-perturbative nature of these types of matter-gravity models. Physically, this motivates why the SM could emerge as a perturbative theory from a gravity-matter model just below the Planck scale. Mathematically, this provides us with a handle to control calculations of matter-gravity models with functional RG techniques.

The crucial question in this context is whether or not gravitational fixed-point values fall into this preferred region of parameter space or not. We point out that two distinct approaches to calculating these fixed-point values, see \cite{Pawlowski:2020qer,Reuter:2012id,Reuter:2019byg} for reviews (which differ because of the breaking of diffeomorphism symmetry by the functional RG setup) are not yet converged. Here, we quote results from background field calculations, which satisfy an (auxiliary) background diffeomorphism symmetry. In these calculations, the gravitational fixed-point value falls into the excluded region under the impact of one and even two generations of SM matter, but moves into the allowed region, once the third generation is added, \cite{Dona:2013qba, Eichhorn:2017ylw}.

\subsection{Higgs mass}
In \cite{Shaposhnikov:2009pv} it was proposed that the Higgs quartic coupling could be predicted, starting from a fixed point with gravity. This would make the ratio of Higgs mass to electroweak scale predictable.
In the meantime, evidence for this mechanism has accumulated \cite{Eichhorn:2017als,Pawlowski:2018ixd,Wetterich:2019zdo,Wetterich:2019rsn,Domenech:2020yjf,Eichhorn:2020sbo,Ohta:2021bkc}. In contrast to the prediction of masses of fermions from asymptotic safety, the numerical value of the prediction is much less uncertain, because the prediction only depends on the sign, but not the size, of the quantum gravitational corrections. Whether the prediction agrees with experiment depends on the top quark mass \cite{Bezrukov:2014ina}; in BSM settings, asymptotic safety could also predict a lower value of the Higgs mass, compatible with observations even at high values of the top quark mass \cite{Eichhorn:2021tsx}.

\subsection{4 dimensions}
What is special about 4 spacetime dimensions in the asymptotic-safety framework? The answer is that the interactions of the SM are marginal in 4 spacetime dimensions. In turn, this is key to achieve a UV completion of the SM through $f_g>0$. In $d>4$, the beta function of the Abelian gauge coupling $g_Y$ takes the form
\be
\beta_{g_Y} = \left(\frac{d-4}{2}-f_g(d) \right)g_Y+...,
\ee 
such that $f_g>(d-4)/2$ is necessary for this sector to be UV complete. In turn $f_g>(d-4)/2$ is achievable, if gravitational fluctuations are strong enough. However, in Sec.~\ref{subsec:WGB}, we discussed that gravitational fluctuations must not become too strong, least they trigger new divergences in higher-order couplings. Thus, $d=4$ (and potentially $d=5$) appear to be the only dimensions, in which $f_g$ can be achieved without exceeding the weak-gravity bound and losing the fixed point \cite{Eichhorn:2019yzm,Eichhorn:2021qet}. 

\section{Beyond the Standard Model}
\subsection{Dark matter models}
A major challenge for the search for dark matter is the huge, theoretically viable parameter space of dark matter models. First, the composition of dark matter (e.g., whether it is one or several species of fields), second, its mass scale and third, its interactions, are theoretically very little constrained. Thus, experiments face a tremendous challenge.\\
Asymptotic safety could provide constraints on dark matter, such that parts of the parameter space may be theoretically excluded due to the predictive power of asymptotic safety.\\
A first example is the Higgs portal coupling $\lambda_H\, H^{\dagger}H\, \phi^2$, between the Higgs field $H$ and a dark scalar $\phi$ that is uncharged. The beta function of that coupling takes the form
\be
\beta_{\lambda_H} =- f_{\lambda}(G) \lambda_H + \mathcal{O}(\lambda_H^2),
\ee
where $f_{\lambda}(G)$ is a function of the dimensionless Newton coupling $G$ and further gravitational couplings, and is negative. Thus, $\lambda_H$ is driven to zero under the impact of gravity fluctuations, i.e., the dark scalar decouples \cite{Eichhorn:2017als}. Demanding asymptotic safety, we find a fixed point at $\lambda_{H\, \ast}=0$, which is infrared attractive. Thus, the portal coupling is zero in the transplanckian regime, where gravity fluctuations prevent its growth. Below the Planck scale, it cannot be generated, because once set to zero at the Planck scale, matter fluctuations cannot generate it. Thus, the ``vanilla" model of dark matter of a single dark scalar is not compatible with asymptotic safety, if it is to be generated as a thermal relic, which needs $\lambda_H\neq 0$.\\
This demonstration of the predictive power of asymptotic safety was followed by the development of models of extended dark sectors which circumvent the decoupling result due to a more intricate structure of the dark sector. \cite{Reichert:2019car} and \cite{Hamada:2020vnf} use additional fields that can regenerate the Higgs portal below the Planck scale. In those cases, asymptotic safety increases the predictive power of the model, compared to an effective field theory setting. A large increase in predictive power was also found in \cite{Eichhorn:2020kca}, in which the dark sector consists of the dark scalar and a dark fermion, which interact through a Yukawa coupling and are coupled to the SM through the Higgs portal. In a toy model of the system, it was found that in the dark sector, the Yukawa coupling, the dark self-interaction, the non-minimal coupling, the mixing angle to the SM scalar and the Higgs portal coupling are all calculable as a function of the dark scalar mass. The model is not viable, if the dark sector does not undergo spontaneous symmetry breaking, such that the dark fermion becomes massive and becomes the dark-matter candidate. Further, the dark scalar mixes with the SM Higgs, lowering the mass of the Higgs within the toy-model study \cite{Eichhorn:2021tsx}. \footnote{At a top mass of 173 GeV, asymptotic safety predicts a Higgs mass a few GeV above the measured value \cite{Shaposhnikov:2009pv}. The top mass is subject to systematic uncertainties, so it is presently unclear whether or not new physics (such as the mixing with a dark scalar) is required to reconcile asymptotic safety and the measured Higgs mass.}

\subsection{Further BSM settings}
Besides settings with dark matter and/or additional cosmological scalar fields, asymptotic safety with gravity has already been explored in the context of the muon magnetic moment \cite{Kowalska:2020zve}, flavor anomalies \cite{Kowalska:2020gie}, axion-like particles \cite{deBrito:2021hde}, Majorana masses \cite{DeBrito:2019rrh}, neutrino masses \cite{Domenech:2020yjf} and grand unified theories \cite{Eichhorn:2017muy,Eichhorn:2019dhg}. In these works, the predictive power of asymptotic safety typically plays a key role in narrowing down the parameter space compared to effective field theory settings. On the one hand, this is quite useful for model building, because it provides a small set of models that are not just phenomenologically relevant, but could also be also fundamentally viable. On the other hand, this is important in order to subject asymptotic safety to further observational tests in the future.

\subsection{Scalar potentials and cosmology}
Upcoming observations that will determine the equation of state of dark energy more precisely could provide valuable information on the underlying quantum-gravity theory \cite{Heisenberg:2018yae}. 
In asymptotic safety, quantum fluctuations of gravity drive scalar potentials \emph{towards} flatness (with the exception of quadratic terms, that typically remain relevant and thus correspond to free parameters in the infrared). As two specific examples of this more general trend, the ratio of Higgs mass to electroweak scale is predicted from a vanishing quartic coupling at the Planck scale, and the decoupling of a single uncharged dark scalar is predicted from a vanishing Higgs-portal coupling at the Planck scale. For potentials in cosmology, this could mean that an arbitrarily flat potential could be accommodated in asymptotic safety \cite{Eichhorn:2020sbo}, as well as a simple cosmological constant \cite{Gubitosi:2018gsl}.

\section{Key open questions on asymptotically safe gravity-matter models}
Key open questions can be divided into questions of a more technical nature and questions of a more conceptual nature.

Open questions of technical nature are questions where proposals for dynamical mechanisms (e.g., for the prediction of certain SM couplings) exist, but the theoretical uncertainties are too large to say with certainty whether or not such a mechanism predicts the correct value. Uncertainties arise on the one hand from the use of truncations, and can be reduced by successively choosing larger truncations. Uncertainties arise on the other hand from the use of Euclidean signature, which, at least in a non-perturbative quantum-gravity regime, does not straightforwardly connect to Lorentzian signature. In this context, it is intriguing that asymptotically safe gravity-matter systems could be near-perturbative. In that case, a regime with a (near-flat) background with small fluctuations about it, could emerge dynamically in asymptotic safety. In turn, such a regime might admit an analytical continuation.\\
A different strategy to reduce systematic uncertainties is by use of another method, e.g., dynamical triangulations, in which some progress on the inclusion of matter degrees of freedom has been made, and early results support previous findings with functional RG techniques. 
One may hope that with the further advancement of these lattice techniques, which are affected by different systematic uncertainties than functional RG studies, complementary evidence for various properties of asymptotically safe gravity-matter systems can be found.

Open question of conceptual nature include, but are not limited to:\\
i) What is the nature of dark matter in asymptotic safety?\\
ii) Is there a preferred mechanism for baryogenesis in asymptotic safety?\\ 
iii) Which mechanisms to generate neutrino masses are compatible with asymptotic safety?\\
iv) Is asymptotic safety compatible with BSM physics that can explain the muon magnetic moment as well as flavor anomalies?\\
These and other questions could on the one hand provide theoretical guidance for particle physics phenomenology and ultimately even experimental searches. On the other hand, the high predictive power that asymptotically safe gravity-matter models appear to have could imply asymptotic safety makes rather specific predictions for the answers to these questions, providing powerful ways to test asymptotic safety through observations.

\acknowledgments
The author acknowledges support by VILLUM FONDEN under grant no.~29405 and would like to thank all her collaborators on gravity-matter systems for many useful discussions.

\bibliographystyle{abbrv}
\bibliography{references}

\end{document}